\documentclass[%
 reprint,
 amsmath,amssymb,
 aps,
]{revtex4-2}

\usepackage{graphicx}
\usepackage{dcolumn}
\usepackage{bm}
\usepackage{array}
\usepackage{mathrsfs}
\usepackage{times}
\usepackage{ragged2e}
\usepackage{enumitem}
\usepackage{etoolbox} 

\newtoggle{issupplement}
\togglefalse{issupplement}

\begin{document}

\preprint{APS/123-QED}

\title{Cosmological Frequency Combs}

\author{Oem Trivedi$^{1}$, Madhurendra Mishra $^{2}$ and Adarsh Ganesan$^{3,4}$}
\affiliation{$^{1}$Department of Physics and Astronomy, Vanderbilt University, Nashville, TN 37235, USA}
\email{Email: oem.trivedi@vanderbilt.edu}
\affiliation{$^{2}$Department of Physics, Sri Guru Tegh Bahadur Khalsa College, University of Delhi, New Delhi, India 110007}
\email{Email: madhurendramishra24@gmail.com}
\affiliation{$^{3}$Department of Electrical and Electronics Engineering, Birla Institute of Technology and Science, Pilani - Dubai Campus, Dubai International Academic City, Dubai, UAE 345055}
\email{Email: adarsh@dubai.bits-pilani.ac.in}
\affiliation{$^{4}$Department of Mechanical Engineering, Birla Institute of Technology and Science, Pilani - Pilani Campus, Vidya Vihar, Pilani, India 333031}

\date{\today}

\begin{abstract}
We identify a new class of time-periodic attractor solutions in scalar–field cosmology, which we term \textit{Cosmological Frequency Combs} (CFC). These solutions arise in exponential quintessence models with a phantom matter background and exhibit coherent phase-locked oscillations in the scalar field's normalized variables. We demonstrate that such dynamics induce modulations in observables like the Hubble parameter and growth rate, offering a dynamical mechanism to even address the $H_0$ tension exactly. Our results uncover a previously unexplored phase of cosmic acceleration, linking the concept of frequency combs to large-scale cosmological evolution.
 
\end{abstract}

\maketitle

\textit{Introduction}-- The finding that Universe's late-time expansion is currently in an accelerating phase marked a pivotal moment in cosmology \cite{SupernovaSearchTeam:1998fmf} and since this discovery, a lot of important efforts have been dedicated to understanding the cause of this expansion. The simplest approach is the cosmological constant $\Lambda$ \cite{Weinberg:1988cp,Padmanabhan:2002ji}, which alongside cold dark matter (CDM) gives us the standard $\Lambda$CDM model.
There are a huge array of more complex proposals which include modified gravity \cite{Nojiri:2010wj,Nojiri:2017ncd} and various models involving scalar fields as the drivers of late-time cosmic acceleration \cite{Zlatev:1998tr,Odintsov:2023weg}. Furthermore, we have potential quantum gravity theories such as braneworld cosmology in string theory, loop quantum cosmology, and asymptotically safe cosmology, which have all proposed their own versions to address late-time acceleration \cite{Sahni:2002dx,Sami:2004xk,Tretyakov:2005en}. There are still very vivid issues though, most notably the Hubble tension, \cite{Planck:2018vyg,riess2019large,riess2021comprehensive} which goes to highlight the limitations of our current cosmological understanding. So, the present cosmic epoch raises profound questions, suggesting that we probably need to take nontrivial approaches to cosmology to fully understand the current phase and future of the universe.

With the complexity of dark energy models increasing, focus started to be given to procedures where one would not have to analytically solve for key cosmological quantities and still be able to find important information about the properties of the expansion. One of the key procedures in this vein emerged as that of employing dynamical systems procedures. Such methods allow one to consider the cosmological system inclusive of the Friedmann, Raychaudhri, continuity equations etc. as a dynamical system of autonomous differential equations written in terms of expansion normalized variables. This allowed one to draw key physical implications based on studying aspects of these systems like fixed points, saddle points, attractor behavior, etc. \cite{dyn1Copeland:2006wr,dyn2bahamonde2018dynamical,dyn3clifton2012modified,dyn4odintsov2018dynamical}. From this the conclusions drawn resulted in the classification of models in various ways, like classifying quintessence models based on thawing, freezing, and scaling classes etc. 

In this work,, we shall consider the dynamical system representation of dark energy and weshallll consider a particularly elementary model, that of quintessence with a simple exponential potential \cite{exp1Barreiro:1999zs,dyn1Copeland:2006wr}. We will demonstrate a novel behavior that is being displayed in this system, termed "Cosmological Frequency Combs" or CFC. Frequency combs represent an array of discrete frequencies separated by a fixed repetition frequency. In the time domain, these combs constitute a train of phase-coherent pulses. While the concept of frequency combs has revolutionized optical frequency metrology and spectroscopy \cite{hall2006nobel,del2007optical,picque2019frequency}, it is only in this decade that their realization has extended to phononic \cite{cao2014phononic,ganesan2017phononic,ganesan2018excitation,nguyen2021spectrally,wu2022hybridized,han2022superconducting,de2023mechanical,jia2025synthesized} and magnonic systems \cite{wang2021magnonic,wang2022twisted,xu2023magnonic,liu2023generation,wang2024enhancement}. Motivated by the recent observations of frequency combs in these condensed matter systems, this work examines the feasibility of such structures manifesting within cosmological frameworks. 

CFC represent a novel class of solutions in scalar-field cosmology where the Universe’s expansion exhibits coherent, self-sustained oscillations rather than monotonic evolution. We show that these arise from limit-cycle attractors in the presence of phantom-like background matter, breaking continuous time-translation symmetry into a discrete structure. CFC offer a new dynamical phase of cosmic acceleration with potential observational signatures in late-time expansion and growth histories. In this letter, we prove the existence of such frequency comb behavior in a basic cosmological model and discusses its concomitant implications.

\textit{Dynamical Systems Framework for Quintessence}--
We start from the canonical single–field action 
\begin{equation}
S=\!\int d^{4}x\sqrt{-g}\left[\frac{R}{2\kappa^{2}}-\frac12 g^{\mu\nu}\partial_{\mu}\phi\partial_{\nu}\phi-V(\phi)\right],
\label{eq:action}
\end{equation}
where $\kappa^{2}=8\pi G$ and $V(\phi)>0$. 
For a spatially flat FLRW metric,
$
ds^{2}=-dt^{2}+a^{2}(t)\bigl(dx^{2}+dy^{2}+dz^{2}\bigr),
$
variation of~\eqref{eq:action} delivers the Friedmann and Klein–Gordon equations
\begin{subequations}
\begin{align}
3H^{2} &= \kappa^{2}\!\left(\rho+\frac{\dot\phi^{2}}{2}+V\right), \label{eq:Friedmann}\\
2\dot H+3H^{2} &= -\kappa^{2}\!\left(w\rho+\frac{\dot\phi^{2}}{2}-V\right),\label{eq:acc}\\
\ddot\phi+3H\dot\phi+V_{\phi}&=0\label{eq:KG}
\end{align}
\end{subequations}
Here $\rho$ is the background barotropic fluid with constant equation-of-state parameter $w$ of matter. Following Copeland et.al. \cite{dyn1Copeland:2006wr}, we introduce the expansion normalized variables
\begin{equation}
x\equiv\frac{\kappa\dot\phi}{\sqrt6\,H}, 
\qquad 
y\equiv\frac{\kappa\sqrt{V}}{\sqrt3\,H},
\label{eq:EN}
\end{equation}
for which the Friedmann constraint reads $1=\Omega_{m}+x^{2}+y^{2}$ with $\Omega_{m}\equiv\kappa^{2}\rho/3H^{2}$.  
The dark–energy density parameter and equation of state become  
$\Omega_{\phi}=x^{2}+y^{2}$ and $w_{\phi}=(x^{2}-y^{2})/(x^{2}+y^{2})$.
\vspace{4pt}
We now choose the potential in exponential form. 
\begin{equation}
V(\phi)=V_{0}e^{-\lambda\kappa\phi},
\qquad 
\lambda=\text{const},
\label{eq:expo}
\end{equation}
This makes the system autonomous, because $\lambda$ is now a parameter, not a dynamical variable. Differentiating \eqref{eq:EN} with respect to $N\equiv\ln a$ and using Eqs.\eqref{eq:Friedmann}–\eqref{eq:KG} one obtains
\begin{subequations}
\begin{align}
x' &= -\frac{3}{2}\!\left[2x+(w-1)x^{3}+x(w+1)(y^{2}-1)
 -\sqrt{\frac{2}{3}}\lambda y^{2}\right], \label{eq:dynx}\\
y' &= -\frac{3}{2}y\!\left[(w-1)x^{2}+(w+1)(y^{2}-1)
 +\sqrt{\frac{2}{3}}\lambda x\right],\label{eq:dyny}
\end{align}
\label{eq:dynsys}%
\end{subequations}
where a prime denotes $d/dN$.  These two first–order equations constitute the compact dynamical system that underpins the analysis in the remainder of this Letter. To probe the emergence of periodic solutions in the scalar-field dynamics, we introduce a harmonic ansatz that captures the possibility of limit-cycle behavior near the non-hyperbolic fixed point of the dynamical system. Motivated by the oscillatory nature of the linearized equations in the presence of a phantom background, we posit that the expansion-normalized variables take the form 
 \begin{equation}
 \begin{aligned}
      x(t) = \frac{1}{2}\left(u\, e^{i\omega N} + u^*\, e^{-i\omega N}\right), \\ \qquad
y(t) = \frac{1}{2}\left(v\, e^{i\omega N/2} + v^*\, e^{-i\omega N/2}\right) 
\end{aligned}
 \end{equation}
where $u(N)$ and $v(N)$ are slowly varying amplitudes, and $\omega$ is the fundamental frequency of the underlying comb structure. This ansatz captures the leading-order time periodic behavior induced by weak nonlinearities near the fixed point and naturally leads to a multiple-timescale decomposition of the dynamics. One finds that the complex envelopes obey the quadratically-coupled equations
\begin{equation}
\begin{aligned}
u^\prime=(\Omega_{u}-i\omega)u+\alpha v^{2}, \\ \qquad
v^\prime=\Bigl(\Omega_{v}-i\frac{\omega}{2}\Bigr)v-\alpha u v^{\ast}
\end{aligned}
\end{equation}

with $\Omega_{u}= \tfrac32(w-1)$, $\Omega_{v}= \tfrac32(w+1)$ and $\alpha=\tfrac12\sqrt{\tfrac32}\lambda$. This pair can be identified as the minimal nonlinear circuit capable of locking phases and generating an infinite ladder of equally spaced spectral lines, which would be a frequency comb in the scalar–field sector.
 
Physically, the harmonic form could be justified by the fact that in the regime where the Jacobian of the system has complex eigenvalues with small real parts, the trajectory executes underdamped oscillations. In this setting, the Hubble drag acts as a weak dissipative force, and the nonlinearity of the scalar field potential provides the restoring and self-interacting structure necessary for stable phase locking. Cosmologically such limit cycles could represent ``time-crystalline'' attractors, where the scalar field drives expansion with discrete time-translation symmetry, modulating the background evolution in a coherent and periodic fashion. 
 
\begin{figure*}
    \centering
    \includegraphics[width=0.8\linewidth]{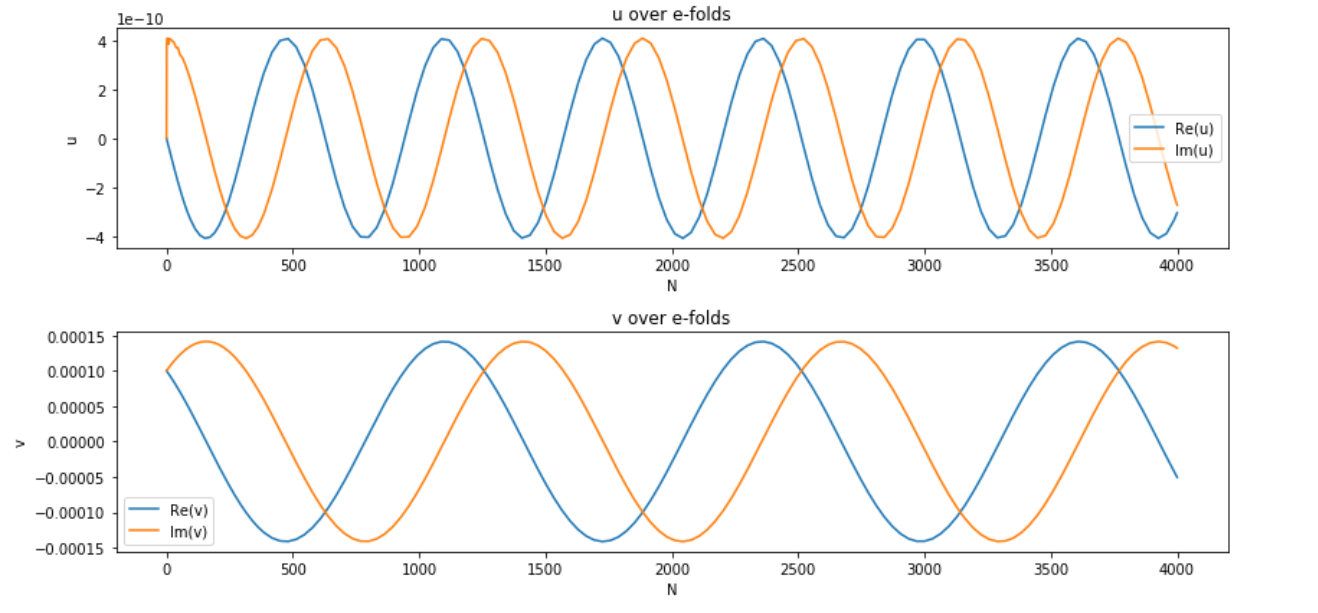}
    \caption{Evolution of the harmonic ansatz amplitudes $u(N)$ and $v(N)$ over cosmic time, expressed as the e-folding number N. The real and imaginary components of each amplitude show coherent oscillations across thousands of e-folds, revealing the phase-locked dynamics responsible for cosmological frequency comb structure.}
    \label{fig1}
\end{figure*}

The validity of the ansatz is confined to regimes where the amplitude of oscillations remains small, $|x|, |y| \ll 1$, and where nonlinear mode coupling is dominated by the fundamental and first subharmonic. Higher harmonics and secular growth are neglected at leading order, and the analysis assumes that the scalar field remains close to the scaling point throughout the cycle. As such, the results are most accurate near the onset of comb behavior and may require numerical validation or higher-order corrections in regimes with large amplitude or steep potentials. In figure \ref{fig1} we have displayed the evolution of the complex amplitudes $u(N)$ and $v(N)$. Their real and imaginary parts oscillate coherently over several thousand e-folds, revealing a deeply phase-locked structure. This extended timescale and large e-folds here are not accidental as it directly results from adopting the small $\omega$ approximation in the CFC analysis, which implies that the modulation frequency per e-fold is extremely low. As a result, each full comb cycle spans hundreds to thousands of e-folds. While such large durations may extend beyond the physically realized history of the universe, they remain meaningful in this model because the observable impact of the comb manifests as slow residual modulation in cosmological quantities like $H(z)$ (which we shall see later on), even within a narrow window of e-folds. 

\begin{figure}
    \centering
    \includegraphics[width=1\linewidth]{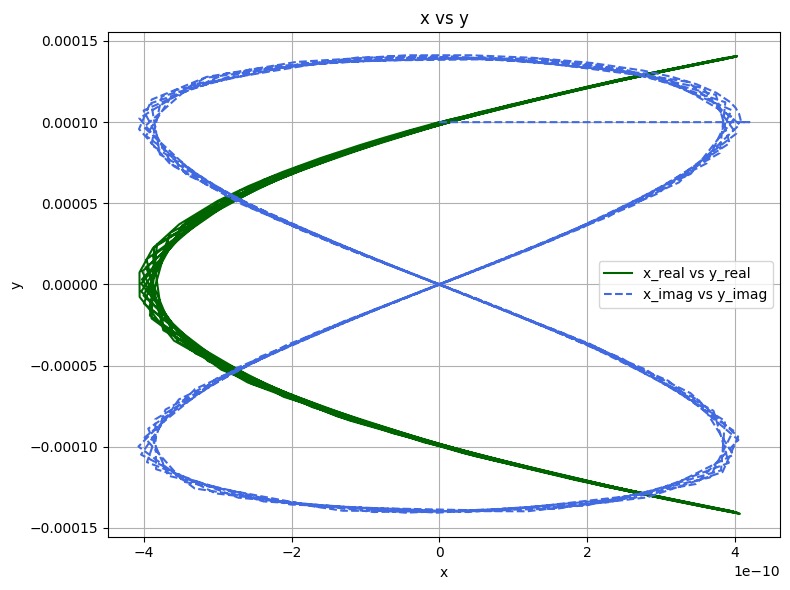}
    \caption{Phase space portrait of the expansion-normalized scalar field variables x and y, showing both real (solid green) and imaginary (dashed blue) components. The figure illustrates the phase-locked, multi frequency oscillations at the heart of cosmological frequency comb dynamics.}
    \label{fig2}
\end{figure}

In figure \ref{fig2}, we have shown the dynamical relationship between the variables x and y, derived from the complex harmonic amplitudes $u(N)$ and $v(N)$. The trajectory traced by $ \Re(x) $ versus $ \Re(y) $ and $ \Im(x) $ versus $ \Im(y) $ reveals a closed, multi loop structure characteristic of multi scale, phase-locked oscillations. This behavior encodes the nonlinear coupling between the kinetic and potential sectors of the scalar field and provides a dynamical systems view of how cosmological frequency combs arise internally. The figure complements the time evolution plots by highlighting how the scalar field’s two dynamical degrees of freedom remain coherently linked over many e-folds, supporting the idea that the comb-like structure is dynamically generated and not externally imposed.  
We would like to emphasize here that while the full comb structure emerges only over a large number of e-folds, in the physically realistic regime of small $N$ (corresponding to redshifts $z \lesssim 2$), the dynamics appear effectively quasi-static. In this regime the phase portraits do not exhibit closed loops, and the oscillatory structure is not yet fully appreciated in the portraits. We use the term "quasi-static" to describe this transitional behavior, distinguishing it from the phase-locked comb regime that arises at late times.

By substituting $ u = u_0 + \delta u\, e^{\eta N}, \quad v = v_0 + \delta v\, e^{\eta N} $ in (7) and considering $\omega \ll \lvert w \rvert$, we obtain \cite{mypaper}
\begin{equation}
\begin{aligned}
\eta^4 - 6 w \eta^3 + \frac{9}{4}(w-1)(9 w + 7) \eta^2 \\ - \frac{27}{4}(w^2 - 1)(11 w + 5) \eta + \frac{405}{4}(w^2 - 1)^2 = 0.
\end{aligned}
\end{equation}

We can now summarize the main cosmological learnings that we get from this analysis as follows:
\\
\\
\textbf{1) A new cosmological realization:}
We have identified a new class of attractor solutions in scalar–field cosmology which we term as Cosmological Frequency Combs, in which the scalar field, evolving under an exponential potential in the presence of a background fluid with equation-of-state $ w < -1 $ settles into a stable nonlinear limit cycle. These time-periodic solutions spontaneously break continuous time-translation symmetry and constitute a new dynamical phase of cosmic acceleration, distinct from de Sitter or monotonic quintessence evolution. The stability region is sharply defined by Routh–Hurwitz conditions leading to the bound $ -21 \leq w \leq -1 $, and arises naturally when the background fluid has a phantom-like equation of state. It is important to mention here, however, that all the properties of the CFC behaviour (phase portraits, observational implications as discussed later etc.) work best when $w \sim -1$, so unrealistic phantom matter EOS are not really favored in our work.
\\
\\
\textbf{2) Possible observational signatures:}
The oscillatory dynamics of the scalar field can induce phase-locked modulations in cosmological observables such as the Hubble parameter and the growth rate of structure. From the expansion-normalized variables \eqref{eq:EN}, the total scalar-field energy density is given by $ \Omega_\phi = x^2 + y^2 $, and since $ x(N) $ and $ y(N) $ are oscillatory due to the harmonic balance ansatz, this implies
\begin{equation}
    \Omega_\phi(N) = \bar{\Omega}_\phi + \delta \cos(\omega N + \varphi)
\end{equation}
which leads to a modulated Hubble parameter of the form
\begin{equation}
    H(N) = \bar{H}(N)\left[1 + \varepsilon \cos(\omega N + \varphi)\right], \quad \varepsilon \sim \mathcal{O}(\lambda)
\end{equation}
This modulation propagates into the structure growth rate through the matter perturbation equation expressed in terms of the e-folding number as
\begin{equation}
    \delta'' + \left(2 + \frac{H'}{H} \right)\delta' - \frac{3}{2} \Omega_m(N)\, \delta = 0
\end{equation}
leading to a periodically varying growth factor
\begin{equation}
    f(N) = \frac{d \ln \delta}{d N} = \bar{f}(N)\left[1 + \varepsilon_f \cos(\omega N + \varphi_f)\right]
\end{equation}
These signatures could be detectable as coherent, low-amplitude residuals in redshift–drift observations, supernova distances, or growth data from next-generation surveys like CMB-S4, ELT and SKA \cite{CMB-S4:2020lpa,ghtbelomestnykh20228,pleabazajian2016cmb,Weltman:2018zrl,Padovani:2023dxc}.
Furthermore, the scalar field's oscillatory dynamics subtly affect the gravitational potential $ \Phi(N) $, which is integrated over the line of sight in weak lensing observables. This leads to a periodic modulation in the convergence power spectrum $ P_\kappa(\ell, N) $, especially at low multipoles and redshifts $ z < 1 $, of the form 
\begin{equation} 
P_\kappa(\ell, N) = \bar{P}_\kappa(\ell, N) \left[1 + \varepsilon_\kappa \cos(\omega N + \varphi)\right]
\end{equation}
Such modulations could be detectable by Euclid \cite{Euclid:2024few} or LSST \cite{LSST:2022kad}, particularly when stacking convergence maps across tomographic redshift bins, and this would provide an independent probe of the CFC phase structure beyond distances and growth.
\\
\\
\textbf{3) Possible addressal of the $H_0$ tension:}
The e-folding–periodic modulation in $H(N)$ provides a potential mechanism for addressing the Hubble tension. Because local measurements of $H_0$ effectively probe the instantaneous value of the Hubble parameter, while CMB constraints measure a time-averaged background \cite{Planck:2018vyg} $\bar{H}$, a phase offset in the frequency comb can lead to an apparent discrepancy:
\begin{equation} \label{h0eqn}
    H_0^{\text{local}} = \bar{H}_0(1 + \varepsilon \cos(\omega N_0 + \varphi)) 
\end{equation}
\begin{figure}
    \centering    \includegraphics[width=1.1\linewidth]{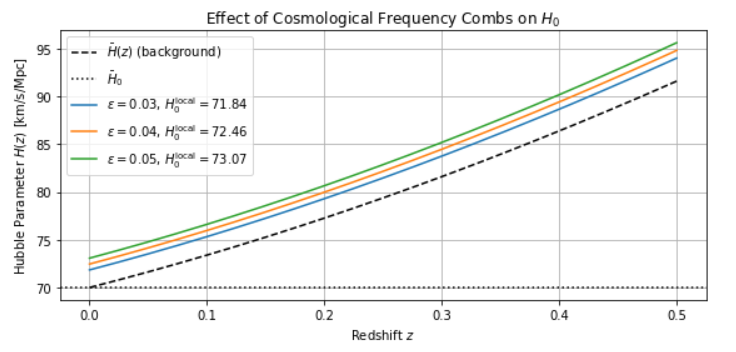}
    \caption{Effect of varying the modulation amplitude $\varepsilon$ in CFC on the Hubble parameter near $z = 0$. Increasing $\varepsilon$ raises the locally inferred Hubble constant $H_0^{\mathrm{local}}$ above the background value $\bar{H}_0$, illustrating a possible dynamical resolution to the Hubble tension.}
    \label{h0tension}
\end{figure}
In figure \ref{h0tension} we show how the locally measured Hubble constant $H_0^{\mathrm{local}}$ is modified in the CFC framework for different values of the modulation amplitude $\varepsilon = 0.03, 0.04, 0.05$. As shown in \eqref{h0eqn}, $\varepsilon$ is directly proportional to the slope parameter $\lambda$ of the scalar field's exponential potential $V(\phi) = V_0 e^{-\lambda \kappa \phi}$. In realistic quintessence models, $\lambda$ is typically $\mathcal{O}(0.1)$, making values of $\varepsilon \sim 0.03\text{–}0.05$ physically admissible. The plot demonstrates how modest values of $\varepsilon$ naturally lift $H_0^{\mathrm{local}}$ by $\sim 2\text{–}3$ km/s/Mpc and alleviate the $H_0$ tension, and in particular at $\varepsilon=0.05$ the tension is completely addressed. Also note that this behavior is the same across a huge range of small $\omega$ (the plot stays almost the same with no drastic $H_0$ value changes for $\omega \sim \mathcal{O}(0.01)$ to even $\omega \sim \mathcal{O}(10^{-12})$), which validates our approximation and shows that small $\omega$ CFC behavior still has observable effects. This offers an alternative explanation without modifying early Universe physics, relying instead on late-time non-equilibrium dynamics in the dark sector \cite{DiValentino:2021izs}.
\\
\\
\textbf{4) Motivations for phantom matter studies:}
The CFC solution is opens the door to a systematic exploration of Floquet phases in cosmology, with applications which may be ranging from ultra-late-time acceleration to structured early-Universe dynamics, and connects the mathematics of harmonic-locking and frequency combs to the physics of cosmic evolution. Future works can look to study such cosmologies in more involved setups, for example by considering non-exponential potentials, couplings of various types, modified gravity theories etc. This result also offers a theoretical motivation for revisiting phantom-matter models, especially in the context of interacting dark sectors or modified gravity scenarios, and provides a new classification of cosmological attractors.
\\
\\
\textit{Conclusions--} In this letter, we have introduced and demonstrated the existence of Cosmological Frequency Combs, a novel dynamical phase in scalar–field cosmology which is characterized by limit-cycle solutions that spontaneously break time-translation symmetry. Arising in simple exponential quintessence models with phantom-like backgrounds, these phase-locked oscillations encode a comb like structure in the evolution of the scalar field and induce residual modulations in key cosmological observables. We have shown that CFCs can offer new insights into late time acceleration, present concrete observational signatures in both distance and growth measures, and can very naturally alleviate the $H_0$ tension through dynamical means. Our results open a new window into non-equilibrium cosmological dynamics, linking the idea of frequency combs to the evolving large-scale structure of the universe. 
\\
\\
\textit{Acknowledgements:}
OT was supported in part by the Vanderbilt Discovery Doctoral Fellowship.


\clearpage
\toggletrue{issupplement}  

\onecolumngrid
\renewcommand{\thesection}{S\arabic{section}}
\renewcommand{\theequation}{S\arabic{equation}}
\setcounter{section}{0}
\setcounter{equation}{0}

\vspace*{1cm}
\begin{center}
  {\LARGE \bf Supplemental Material}\\[0.5em]
  {\large Cosmological Frequency Combs}\\[1.5em]
  {\normalsize Oem Trivedi\textsuperscript{*1}, Madhurendra Mishra\textsuperscript{†2}, and Adarsh Ganesan\textsuperscript{‡3,4}}\\[1em]
  {\footnotesize
  \textsuperscript{1}Department of Physics and Astronomy, Vanderbilt University, Nashville, TN 37235, USA \\
  \textsuperscript{2}Department of Physics, Sri Guru Tegh Bahadur Khalsa College, University of Delhi, New Delhi, India 110007 \\
  \textsuperscript{3}Department of Electrical and Electronics Engineering, BITS Pilani - Dubai Campus, UAE 345055 \\
  \textsuperscript{4}Department of Mechanical Engineering, BITS Pilani - Pilani Campus, Vidya Vihar, Pilani, India 333031
  }
\end{center}
\vspace{1cm}

\justify A quintessence scalar field paradigm given by the action \begin{equation*}
S=\!\int d^{4}x\sqrt{-g}\left[\frac{R}{2\kappa^{2}}-\frac12 g^{\mu\nu}\partial_{\mu}\phi\partial_{\nu}\phi-V(\phi)\right],
\end{equation*}
is governed by the following cosmological equations 
\begin{align*}
3H^{2} &= \kappa^{2}\!\left(\rho+\frac{\dot\phi^{2}}{2}+V\right), \\
2\dot H+3H^{2} &= -\kappa^{2}\!\left(w\rho+\frac{\dot\phi^{2}}{2}-V\right),\\
\ddot\phi+3H\dot\phi+V_{\phi}&=0
\end{align*}
Here $\rho$ is the background barotropic fluid with constant equation-of-state parameter $w$ of matter. Defining expansion normalized variables as in \cite{dyn1Copeland:2006wr,dyn2bahamonde2018dynamical,dyn3clifton2012modified}, \begin{equation*}
x\equiv\frac{\kappa\dot\phi}{\sqrt6\,H}, 
\qquad 
y\equiv\frac{\kappa\sqrt{V}}{\sqrt3\,H},
\end{equation*}
Alongside choosing an exponential potential form
\begin{equation*}
V(\phi)=V_{0}e^{-\lambda\kappa\phi},
\qquad 
\lambda=\text{const},
\end{equation*}
Allows us to write the cosmological equations in the following form
\justify
\begin{align}
x^\prime &= -\frac{3}{2} \left[ 2 x + (w-1) x^3 + x (w+1) (y^2 - 1) - \sqrt{\frac{2}{3}} \lambda y^2 \right], \\
y^\prime &= -\frac{3}{2} y \left[ (w-1) x^2 + (w+1) (y^2 - 1) + \sqrt{\frac{2}{3}} \lambda x \right].
\end{align}
\justify
Rewriting equations (S1) and (S2) as:
\justify
\begin{align}
x^\prime &= \frac{3}{2} (w-1) \left( x - x^3 \right) + \sqrt{\frac{3}{2}} y^2 \left( \lambda - x \sqrt{\frac{3}{2}} (w+1) \right), \\
y^\prime &= \frac{3}{2} (w+1) \left( y - y^3 \right) - \sqrt{\frac{3}{2}} x y \left( \lambda + x \sqrt{\frac{3}{2}} (w-1) \right).
\end{align}

\justify
Assuming solutions of the form:

\justify
\begin{align}
x &= \frac{1}{2} \left( u e^{i \omega N} + u^* e^{-i \omega N} \right), \\
y &= \frac{1}{2} \left( v e^{i \frac{\omega}{2} N} + v^* e^{-i \frac{\omega}{2} N} \right).
\end{align}

\justify
The derivatives and products are:

\justify

\begin{align}
x^\prime &= \frac{1}{2} \left( i \omega u e^{i \omega N} + u^\prime e^{i \omega N} - i \omega u^* e^{-i \omega N} + {u^\prime}^* e^{-i \omega N} \right), \\
y^\prime &= \frac{1}{2} \left( i \frac{\omega}{2} v e^{i \frac{\omega}{2} N} + v^\prime e^{i \frac{\omega}{2} N} - i \frac{\omega}{2} v^* e^{-i \frac{\omega}{2} N} + {v^\prime}^* e^{-i \frac{\omega}{2} N} \right), \\
y^2 &= \frac{1}{4} \left( v^2 e^{i \omega N} + v^{*2} e^{-i \omega N} + 2 |v|^2 \right), \\
x y &= \frac{1}{4} \left( u v e^{3 i \frac{\omega}{2} N} + u v^* e^{i \frac{\omega}{2} N} + u^* v e^{-i \frac{\omega}{2} N} + u^* v^* e^{-3 i \frac{\omega}{2} N} \right), \\
y^2 x &= \frac{1}{8} \left( u v^2 e^{2 i \omega N} + |v|^2 u e^{i \omega N} + u v^{*2} + u^* v^2 + |v|^2 u^* e^{-i \omega N} + u^* v^{*2} e^{-2 i \omega N} \right), \\
x^2 y &= \frac{1}{8} \left( u^2 v e^{5 i \frac{\omega}{2} N} + u^2 v^* e^{3 i \frac{\omega}{2} N} + 2 |u|^2 v e^{i \frac{\omega}{2} N} + 2 |u|^2 v^* e^{-i \frac{\omega}{2} N} + u^{*2} v e^{-3 i \frac{\omega}{2} N} + u^{*2} v^* e^{-5 i \frac{\omega}{2} N} \right), \\
x^3 &= \frac{1}{8} \left( u^3 e^{3 i \omega N} + 3 |u|^2 u e^{i \omega N} + 3 |u|^2 u^* e^{-i \omega N} + u^{*3} e^{-3 i \omega N} \right), \\
y^3 &= \frac{1}{8} \left( v^3 e^{3 i \frac{\omega}{2} N} + 3 |v|^2 v e^{i \frac{\omega}{2} N} + 3 |v|^2 v^* e^{-i \frac{\omega}{2} N} + v^{*3} e^{-3 i \frac{\omega}{2} N} \right).
\end{align}

\justify
Substituting equations (S5) to (S14) into (S3) and (S4) yields:

\justify

\begin{align}
(i \omega u + u^\prime) &= \left[ \frac{3}{2} w \left( 1 - \frac{3}{4} |u|^2 - \frac{1}{4} |v|^2 \right) - \frac{3}{2} \left( 1 - \frac{3}{4} |u|^2 + \frac{1}{4} |v|^2 \right) \right] u + \frac{1}{2} \sqrt{\frac{3}{2}} \lambda (v^2), \\
\left( i \frac{\omega}{2} v + v^\prime \right) &= \left[ \frac{3}{2} w \left( 1 - \frac{3}{4} |v|^2 - \frac{1}{2} |u|^2 \right) + \frac{3}{2} \left( 1 - \frac{3}{4} |v|^2 + \frac{1}{2} |u|^2 \right) \right] v - \frac{1}{2} \sqrt{\frac{3}{2}} \lambda u v^*.
\end{align}

\justify
For small $|u|^2, |v|^2 \ll 1$, equations (S15) and (S16) simplify to:

\justify
\begin{align}
(i \omega u + u^\prime) &= \frac{3}{2} (w-1) u + \frac{1}{2} \sqrt{\frac{3}{2}} \lambda (v^2), \\
\left( i \frac{\omega}{2} v + v^\prime \right) &= \frac{3}{2} (w+1) v - \frac{1}{2} \sqrt{\frac{3}{2}} \lambda u v^*.
\end{align}

\justify
Defining $\Omega_u = \frac{3}{2}(w-1)$, $\Omega_v = \frac{3}{2}(w+1)$, and $\alpha = \frac{1}{2} \sqrt{\frac{3}{2}} \lambda$, we obtain:

\begin{align}
u^\prime &= (\Omega_u - i \omega) u + \alpha v^2, \\
v^\prime &= \left( \Omega_v - i \frac{\omega}{2} \right) v - \alpha u v^*.
\end{align}

\justify
For frequency combs, express solutions as $u = u_0 + \delta u$ and $v = v_0 + \delta v$. Setting $u^\prime = 0$ and $v^\prime = 0$ in equations (S19) and (S20) for steady-state components:

\justify
\begin{align}
(\Omega_u - i \omega) u_0 + \alpha (v_0^2) &= 0, \\
\left( \Omega_v - i \frac{\omega}{2} \right) v_0 - \alpha u_0 v_0^* &= 0.
\end{align}

\justify
Multiplying equations (S21) and (S22) by their complex conjugates:

\justify
\begin{align}
u_0 &= -\frac{\alpha v_0^2}{\Omega_u - i \omega}, \\
u_0^* &= -\frac{\alpha v_0^{*2}}{\Omega_u + i \omega}, \\
|u_0|^2 &= \frac{\alpha^2 |v_0|^4}{\Omega_u^2 + \omega^2}, \\
\left( \Omega_v - i \frac{\omega}{2} \right) v_0 &= \alpha u_0 v_0^*, \\
\left( \Omega_v + i \frac{\omega}{2} \right) v_0^* &= \alpha u_0^* v_0, \\
|u_0|^2 &= \frac{1}{\alpha^2} \left( \Omega_v^2 + \frac{\omega^2}{4} \right).
\end{align}

\justify
Substituting equation (S28) into (S25):

\justify
\begin{equation}
|v_0|^2 = \frac{1}{\alpha^2} \sqrt{\left( \Omega_v^2 + \frac{\omega^2}{4} \right) (\Omega_u^2 + \omega^2)}.
\end{equation}

\justify
Substituting $u = u_0 + \delta u$ and $v = v_0 + \delta v$ into equations (19) and (20):

\justify
\begin{align}
\delta u^\prime &= (\Omega_u - i \omega) u_0 + (\Omega_u - i \omega) \delta u + \alpha v_0^2 + 2 \alpha v_0 \delta v + \alpha \delta v^2, \\
\delta v^\prime &= \left( \Omega_v - i \frac{\omega}{2} \right) v_0 + \left( \Omega_v - i \frac{\omega}{2} \right) \delta v - \alpha u_0 v_0^* - \alpha u_0 \delta v^* - \alpha \delta u v_0^* - \alpha \delta u \delta v^*.
\end{align}

\justify
Using steady-state conditions, simplify to:
\justify
\begin{align}
\delta u^\prime &= (\Omega_u - i \omega) \delta u + 2 \alpha v_0 \delta v + \alpha \delta v^2, \\
\delta v^\prime &= \left( \Omega_v - i \frac{\omega}{2} \right) \delta v - \alpha u_0 \delta v^* - \alpha \delta u v_0^* - \alpha \delta u \delta v^*.
\end{align}

\justify
Neglecting higher-order terms $\alpha \delta v^2$ and $\alpha \delta u \delta v^*$:
\justify

\begin{align}
\delta u^\prime &= (\Omega_u - i \omega) \delta u + 2 \alpha v_0 \delta v, \\
\delta v^\prime &= \left( \Omega_v - i \frac{\omega}{2} \right) \delta v - \alpha u_0 \delta v^* - \alpha \delta u v_0^*.
\end{align}
\justify

Conjugates of equations (S34) and (S35):
\justify

\begin{align}
\delta {u^\prime}^* &= (\Omega_u + i \omega) \delta u^* + 2 \alpha v_0^* \delta v^*, \\
\delta {v^\prime}^* &= \left( \Omega_v + i \frac{\omega}{2} \right) \delta v^* - \alpha u_0^* \delta v - \alpha \delta u^* v_0.
\end{align}
\justify

Assuming $\delta u = b_1 e^{\eta t}$, $\delta u^* = b_2 e^{\eta t}$, $\delta v = b_3 e^{\eta t}$, and $\delta v^* = b_4 e^{\eta t}$, and substituting:
\justify
\begin{align}
b_1 &= \frac{2 \alpha v_0}{\eta - \Omega_u + i \omega} b_3, \\
b_2 &= \frac{2 \alpha v_0^*}{\eta - \Omega_u - i \omega} b_4, \\
b_3 &= -\frac{\alpha u_0 (\eta - \Omega_u + i \omega)}{(\eta - \Omega_v + i \frac{\omega}{2})(\eta - \Omega_u + i \omega) + 2 \alpha^2 |v_0|^2} b_4.
\end{align}
\justify

Substituting equations (S38) to (S40) into (S37):
\justify
\begin{align}
\eta &= \left( \Omega_v + i \frac{\omega}{2} \right) + \frac{\alpha^2 |u_0|^2 (\eta - \Omega_u + i \omega)}{(\eta - \Omega_v + i \frac{\omega}{2})(\eta - \Omega_u + i \omega) + 2 \alpha^2 |v_0|^2} - \frac{2 \alpha^2 |v_0|^2}{\eta - \Omega_u - i \omega}, \\
\eta^2 &- (\Omega_u + \Omega_v) \eta + \Omega_u \Omega_v - \frac{\omega^2}{2} + 2 \alpha^2 |v_0|^2 - i \omega \left( \frac{3}{2} \eta - \frac{\Omega_u}{2} - \Omega_v \right) \notag \\
&= \frac{\alpha^2 |u_0|^2 (\eta^2 - 2 \Omega_u \eta + \Omega_u^2 + \omega^2)}{\alpha^2 (\Omega_v + \Omega_u) \eta + \Omega_u \Omega_v - \frac{\omega^2}{2} + 2 \alpha^2 |v_0|^2 + i \omega \left( \frac{3}{2} \eta - \frac{\Omega_u}{2} - \Omega_v \right)}.
\end{align}

\justify
This leads to the polynomial equation after simplification:
\justify
\begin{align}
&\left[ \left( \eta^2 - (\Omega_u + \Omega_v) \eta + \Omega_u \Omega_v - \frac{\omega^2}{2} + 2 \alpha^2 |v_0|^2 \right)^2 + \omega^2 \left( \frac{3}{2} \eta - \frac{\Omega_u}{2} - \Omega_v \right)^2 \right] \notag \\
&= \alpha^2 |u_0|^2 (\eta^2 - 2 \Omega_u \eta + \Omega_u^2 + \omega^2).
\end{align}
\justify

Substituting $|u_0|^2$ and $|v_0|^2$:
\justify
\begin{align}
&\left[ (\Omega_u + \Omega_v) \eta^3 + \left( 4 \Omega_u \Omega_v + \Omega_u^2 + \omega^2 + 4 \sqrt{\left( \Omega_v^2 + \frac{\omega^2}{4} \right)(\Omega_u^2 + \omega^2)} \right) \eta^2 \right. \notag \\
&\left. + \left( 2 \Omega_u^2 \Omega_v + 2 \Omega_v \omega^2 + 4 \sqrt{\left( \Omega_v^2 + \frac{\omega^2}{4} \right)(\Omega_u^2 + \omega^2)} (\Omega_u + 4 \Omega_v) \right) \eta \right. \notag \\
&\left. + \left( \Omega_v^2 + \frac{\omega^2}{4} \right)(\Omega_u^2 + \omega^2) + 2 \sqrt{\left( \Omega_v^2 + \frac{\omega}{4} \right)(\Omega_u^2 + \omega^2)} (\Omega_u + \omega^2) \right] = 0.
\end{align}
\justify
With $\Omega_u = \frac{3}{2}(w-1)$ and $\Omega_v = \frac{3}{2}(w+1)$:
\justify

\begin{align}
&\eta^4 - 6 w \eta^3 + \left( \frac{9}{4}(w-1)(9 w + 7) + \frac{9}{4} \sqrt{\left( \frac{9}{4}(w+1)^2 + \frac{\omega^2}{4} \right) \left( \frac{9}{4}(w-1)^2 + \omega^2 \right)} \right) \eta^2 \notag \\
&- \left( \frac{27}{4}(w-1)^2 (w+1) + 3 (w+1) \omega^2 + \frac{9}{4} \sqrt{\left( \frac{9}{4}(w+1)^2 + \frac{\omega^2}{4} \right) \left( \frac{9}{4}(w-1)^2 + \omega^2 \right)} (15 w + 9) \right) \eta \notag \\
&+ \frac{9}{4} \left( \frac{9}{4}(w+1)^2 + \frac{\omega^2}{4} \right) \left( \frac{9}{4}(w-1)^2 + \omega^2 \right) \notag \\
&+ \frac{9}{4} \sqrt{\left( \frac{9}{4}(w+1)^2 + \frac{\omega^2}{4} \right) \left( \frac{9}{4}(w-1)^2 + \omega^2 \right)} \left( \frac{9}{4}(w^2 - 1) - \omega^2 \right) = 0.
\end{align}

\justify
Assuming $\omega^2 \ll |w|$:
\justify
\begin{equation}
\eta^4 - 6 w \eta^3 + \frac{9}{4}(w-1)(9 w + 7) \eta^2 - \frac{27}{4}(w^2 - 1)(11 w + 5) \eta + \frac{405}{4}(w^2 - 1)^2 = 0.
\end{equation}
\justify

For stability, apply the Routh-Hurwitz criteria for $\eta^4 + c_1 \eta^3 + c_2 \eta^2 + c_3 \eta + c_4 = 0$, where:

\justify
\begin{itemize}
\item $c_1 = -6 w$,
\item $c_2 = \frac{9}{4}(w-1)(9 w + 7)$,
\item $c_3 = -\frac{27}{4}(w^2 - 1)(11 w + 5)$,
\item $c_4 = \frac{405}{4}(w^2 - 1)^2$.
\end{itemize}

\justify
The conditions are:
\justify

\begin{enumerate}
\item $c_1 \geq 0 \implies -6 w \geq 0 \implies w \leq 0$.
\item $c_2 \geq 0 \implies \frac{9}{4}(w-1)(9 w + 7) \geq 0 \implies w \in (-\infty, -\frac{7}{9}] \cup [1, \infty) \implies w \in (-\infty, -0.778] \cup [1, \infty)$.
\item $c_3 \geq 0 \implies -\frac{27}{4}(w^2 - 1)(11 w + 5) \geq 0 \implies w \in (-\infty, -1] \cup [-\frac{5}{11}, 1] \implies w \in (-\infty, -1] \cup [-0.455, 1]$.
\item $c_4 \geq 0$ is always satisfied.
\item $c_1 c_2 - c_3 \geq 0$:

\begin{align}
&-6 w \cdot \frac{9}{4}(w-1)(9 w + 7) + \frac{27}{4}(w^2 - 1)(11 w + 5) \geq 0, \\
&\implies -7 w^3 + 9 w^2 + 3 w - 5 \geq 0, \\
&\implies (w-1)^2 (7 w + 5) \leq 0, \\
&\implies w \leq -\frac{5}{7} \implies w \leq -0.714.
\end{align}

\item $c_3 (c_1 c_2 - c_3) - c_1^2 c_4 \geq 0$:

\justify
\begin{align}
&\left( \frac{27}{4} \right)^2 (w^2 - 1)(11 w + 5)(w-1)^2 (7 w + 5) - (-6 w)^2 \frac{405}{4}(w^2 - 1)^2 \geq 0, \\
&\implies (w-1)^2 (w+1) (3 w^3 + 67 w^2 + 65 w + 25) \leq 0, \\
&\implies -21.3361 \leq w \leq -1.
\end{align}

\end{enumerate}

Compiling all conditions:
\justify
\begin{itemize}
\item $w \leq 0$,
\item $w \in (-\infty, -0.778] \cup [1, \infty)$,
\item $w \in (-\infty, -1] \cup [-0.455, 1]$,
\item $w \leq -0.714$,
\item $-21.3361 \leq w \leq -1$.
\end{itemize}
\justify
So we see that the region which satisfies all the creation is $-21.3361 \leq w \leq -1$, which suggests that a frequency comb behaviour can be seen in an expanding cosmology if there is a phantom-like matter component.

\bibliography{apssamp}

\end{document}